\documentclass[osajnl,twocolumn,showpacs,superscriptaddress,10pt]{revtex4-1} 
\usepackage{amsmath,amssymb,graphicx}

\begin{document}

\title{Physical key-protected one-time pad}

\author{Roarke Horstmeyer}\email{Corresponding author: roarke@caltech.edu}
\author{Benjamin Judkewitz}
\affiliation{Departments of Electrical Engineering and Bioengineering, California Institute of Technology, Pasadena, CA 91125}

\author{Ivo Vellekoop}
\affiliation{Biomedical Photonic Imaging Group, MIRA Institute for Biomedical Technology and Technical Medicine, University of Twente, P.O. Box 217, 7500 AE Enschede, the Netherlands}

\author{Sid Assawaworrarit}
\author{Changhuei Yang}
\affiliation{Departments of Electrical Engineering and Bioengineering, California Institute of Technology, Pasadena, CA 91125}

%

\maketitle 


One-time pads are commonly acknowledged as the holy grail of cryptography~\cite{1}, but have limited application in modern ciphers. In practice, an unbreakable one-time pad (OTP) protocol requires storage of a large and random key that must remain absolutely safe against malicious attempts to copy it. As demonstrated by numerous recent database breaches, key storage is inherently vulnerable when digital electronic memory is used~\cite{2,3}. In this work, we present a new optical system and communication protocol that allows two parties to securely share gigabits of ideally random OTP keys without digitally saving any sensitive key information. Each key is derived by optically probing the randomness contained within a uniquely complex physical structure and keys are shared with theoretically perfect security. This novel method is extraordinarily resilient to malicious duplication attempts and is capable of securely storing communication keys at an unprecedented density. Our scheme may additionally extend to public key-based protocols, which, as photonic devices begin to solve an increasing number of integrated circuit bottlenecks, indicates volumetric scattering as a natural and efficient communication key database. 

\begin{figure}[t!]
\centerline{\includegraphics[width=.99\columnwidth]{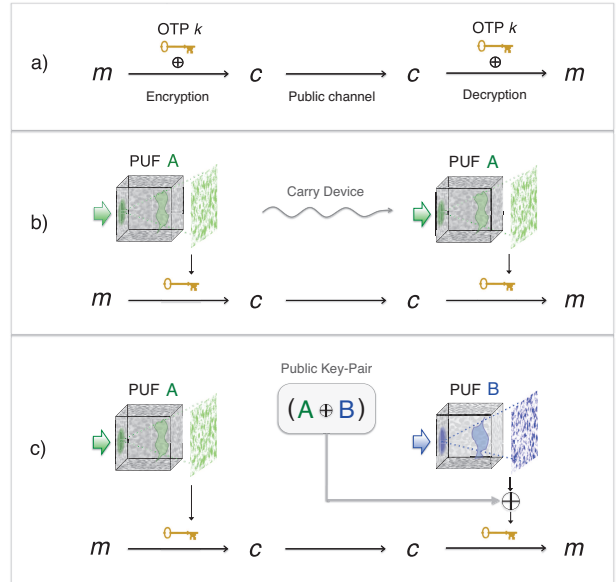}}
\vspace{-.2cm}
\caption{\textbf{The physical OTP. }(a) The theoretically perfect OTP mixes a binary message $m$ with a random digital key $k$ of equal length to create an ideally encrypted ciphertext $c$. (b) Storing keys within a volumetric scatterer's random structure (PUF $A$) helps address the inherent insecurity of digital memory. Keys are accessed with specifically shaped optical probes (green light). If the volumetric scatterer's structure is truly unique and unclonable, then physical transportation appears the only method of sharing keys, which is impractical. (c) The physically secure OTP protocol. The digital XOR of keys from PUF $A$ and $B$, which itself forms an encrypted OTP ciphertext, helps form an information-theoretically secure link between the two uncopyable devices.}
\label{fig1}
\end{figure}

Prior optical methods of establishing encrypted two-party communication include classical spatial~\cite{4} and temporal~\cite{5,6,7,8,9} setups, as well as quantum key distribution~\cite{10} (QKD). Each of these approaches, including the unconditionally secure connection offered by QKD, must electronically save its keys~\cite{11,12}. In a world that increasingly requires secure mobile connectivity, a non-electronic and portable key storage medium resilient against invasive threats~\cite{2,3,13} can eliminate many of conventional electronic memory's intrinsic vulnerabilities~\cite{14}. For example, attacks that can discretely reset~\cite{15}, image~\cite{16} or freeze~\cite{17} the contents of flash memory are well recognized by the security community. Attempted safeguards that apply the inherent randomness within an integrated circuit~\cite{18,19,20}, RFID chip~\cite{21}, and optical scattering medium~\cite{22,23,24,25} implement what is often referred to as a physical unclonable function (PUF). However, due to their significantly limited bit capacity, previous PUF setups have only been demonstrated with terminal-based identification~\cite{22,23,24,25} and public-key protocols~\cite{26,27}. Their potential application to perfectly secure OTP communication is further compounded by the fact that outputs of different PUF devices are independent, and thus not synchronized.

\begin{figure*}[!ht]
\centerline{\includegraphics[width=1.8\columnwidth]{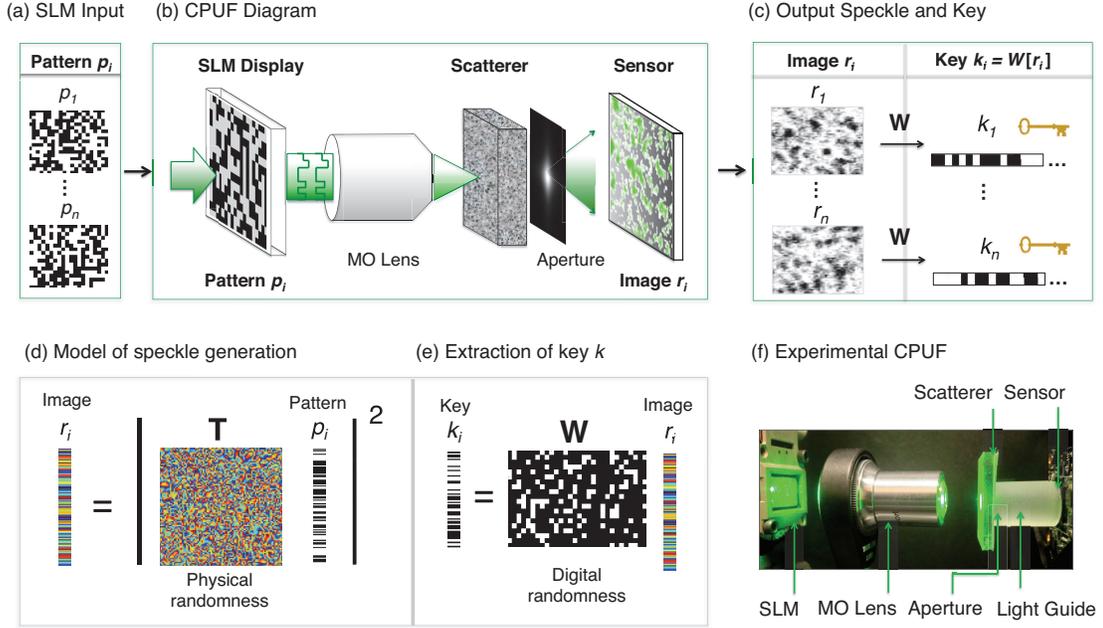}}
\caption{\textbf{The construction and operation of a CPUF.} (a) Sequentially over time, $n$ random phase patterns $p_i$ are displayed on an SLM. (b) A microscope objective (MO) focuses each random wavefront from the SLM onto a volumetric scatterer. The scrambled light emerging from the material passes through a designed aperture before being detected by a CMOS sensor. (c) Each detected speckle image $r$ is digitally transformed into an ideally random key $k$ with a constant digital whitening projection operator $W$. (d) Optical scattering is mathematically represented by a complex random Gaussian matrix $T$ and (e) digital whitening is described by a sparse binary random matrix $W$. The combination of one unique $T$ and general $W$ per CPUF device leads to an ideally random multi-gigabit key space that is very difficult to characterize or clone. (f) The experimental CPUF setup, including all components in (b).}
\label{fig2}
\end{figure*}

 In this paper, we report a novel approach for implementing an optical scattering-based communication PUF (CPUF) that enables the reproducible generation of gigabits of statistically random keys. We further report a novel synchronization protocol, which in combination with a pair of these CPUF devices allows establishment of a physically secured OTP communication link. 

Before delving into specifics, we first briefly outline the operation and security requirements of a successful CPUF link (see Fig.~\ref{fig1}). To communicate, two users, Alice and Bob, will first connect their personal CPUF devices over a known secure connection (e.g., by physically meeting, or by using QKD) to generate a shared random key. Once separate and mobile, they may securely exchange messages over any public channel until all shared key bits are exhausted. Besides offering OTP-strong encryption (i.e., eavesdropping is theoretically impossible), a secure CPUF link must meet the following requirements: first, its security must not depend upon any electronically stored data. Second, a malicious third party (Eve) with temporary access to a CPUF must not be able to efficiently copy or model its contents. And third, if Eve steals a device she must not be able to effectively send or receive messages.

The above requirements are met by storing random keys within the optical device outlined in Fig.~\ref{fig2}. To generate a key, we first illuminate a volumetric scattering medium with a random coherent optical wavefront defined by a spatial light modulator (SLM) (Fig.~\ref{fig2}a). An output field emerges with a profile that depends on both the random input wavefront and the medium's random distribution and orientation of scattering particles. A designed aperture mask then shapes the output field before it propagates to an attached CMOS sensor. The mask is patterned to ensure the output speckle follows Markov statistics -- an important condition for effective random key generation~\cite{28}. A combinatorially large space of possible wavelength-scale interactions enables detection of many mutually random speckle field outputs from a set of different SLM phase profile inputs. We demonstrate how over 10 gigabits of randomness may be optically extracted from within a 2 mm$^{3}$ volume. 

This large amount of extractable physical randomness is described by mathematically representing optical scattering with a random transmission matrix~\cite{29} $T$ (Fig.~\ref{fig2}d). The output scattered field created by displaying the $i^{th}$ random SLM phase pattern $p_i$ may be described by $u_i = T \cdot p_i$. After the sensor detects the output field's intensity $r_i = |u_{i}|^2$, a fixed whitening and noise removal operation $W$ turns the speckle pattern into a verifiably random and repeatable key $k_i$ (Fig. 2e). 

\begin{figure*}[t!]
\centerline{\includegraphics[width=1.55\columnwidth]{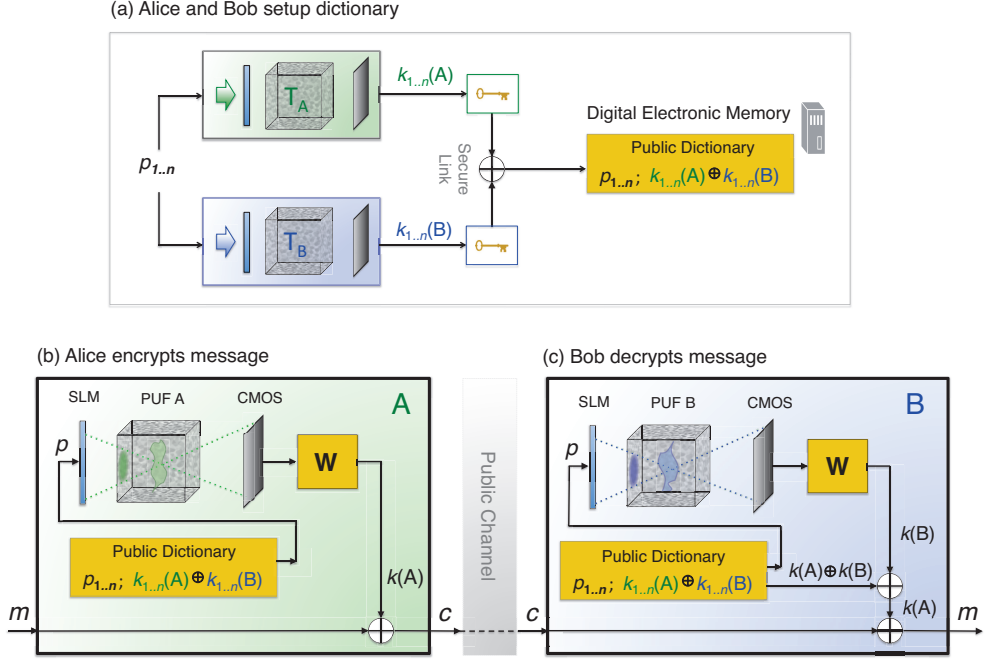}}
\caption{\textbf{Ideally secure CPUF communication protocol and its experimental demonstration.} (a) During setup, Alice and Bob securely connect their two devices and generate $n$ CPUF keys $k_{1..n}(A)$ and $k_{1..n}(B)$ using the same $n$ input SLM patterns $p_{1..n}$. Each key-mixture $k_{i}(A) \oplus k_{i}(B)$ is saved in a digital electronic dictionary that is assumed public, along with its corresponding SLM pattern $p_{i}$. (b) At a later time $t_{c}$, Alice may send Bob a perfectly secure ciphertext $c$ by selecting a pattern $p$, re-creating key $k(A)$, and then XORing this key with her message $m$. The public dictionary can be saved locally on each device without any sacrifice to security. (c) Bob decrypts the received ciphertext. He uses $p$ to both re-generate key $k(B)$ and to find the public dictionary's corresponding key-mixture. The XOR of $c$ with $k(B)$ and the key-mixture reveals $m$. No secret key is ever digitally stored, obfuscating any copy attack. }
\label{fig3}
\end{figure*}

The SLM pattern $p_i$ and output key $k_i$ are thus connected by,
\begin{equation}
k_{i} = W \cdot |T \cdot p_{i} |^{2}.
\end{equation}
The projection of optical field $p_{i}$ into the random matrix $T$, uniquely defined by the scatterer within each CPUF device, imparts key $k_i$ with its unclonable security.

A set of n random keys $k_{1..n}(A)$ generated by Alice's CPUF, along with a corresponding key set $k_{1..n}(B)$ generated by Bob's CPUF, enable physically secure OTP communication with the assistance of a digitally saved public dictionary (Fig.~\ref{fig3}). As described above, Alice and Bob begin by establishing a secure connection between their two devices. While connected, they sequentially probe their scatterers with the same set of $n$ random SLM phase patterns $p_{1..n}$, respectively detecting key sets $k_{1..n}(A)$ and $k_{1..n}(B)$ following equation (1) (Fig.~\ref{fig3}a). Key sets $k_{1..n}(A)$ and $k_{1..n}(B)$ reflect each device's unique transmission matrix $T_A$ and $T_B$, but remain synchronized through Alice and Bob's shared use of SLM set $p_{1..n}$. Without leaving any digital trace of an individual key, Alice and Bob populate a public dictionary with each SLM pattern $p_i$ paired with the digital XOR of the two keys it generates, $k_{i}(A) \oplus k_{i}(B)$, for $1 \le i \le n$. An eavesdropper will gain no information about an individual key from this saved XOR ``key-mixture", since it takes the form of a secure OTP ciphertext. 

Once mobile at a later time $t_c$, Alice may securely send Bob a message $m$ by first randomly selecting an unused pattern $p_i$ from the public dictionary to re-create key $k_{i}(A)$ (Fig.~\ref{fig3}b). Then, Alice may use this key to create and send an XOR-encrypted ciphertext $c$, where $c = k_{i}(A) \oplus m$ (here we assume $k_{i}(A)$ and $m$ are the same length -- longer messages are encrypted by concatenating multiple keys). To complete the protocol, Alice must also send Bob the index $i$ of the SLM pattern $p_{i}$ she displayed, which need not be encrypted.

Bob decrypts Alice's ciphertext using both his CPUF and the public dictionary (Fig.~\ref{fig3}c). He displays $p_{i}$ to optically regenerate key $k_{i}(B)$, and accesses dictionary entry $i$ to obtain the corresponding key-mixture $\left[k_{i}(A) \oplus k_{i}(B)\right]$. The decoded message is then obtained by an XOR of these two sequences with the received ciphertext:
\begin{equation}
k_{i}(B) \oplus \left[k_{i}(A) \oplus k_{i}(B)\right] \oplus \left(k_{i}(A) \oplus m \right) = m
\end{equation}
The total number of secure bits $N$ that Alice and Bob may share is proportional to the product of the number of saved key-mixtures $n$ and the number of bits within each key $|k|$. Factors that limit $N$ include display and sensor resolution, scatterer size, and allowed setup time (Supplement A).

\begin{figure*}[t!]
\centerline{\includegraphics[width=1.7\columnwidth]{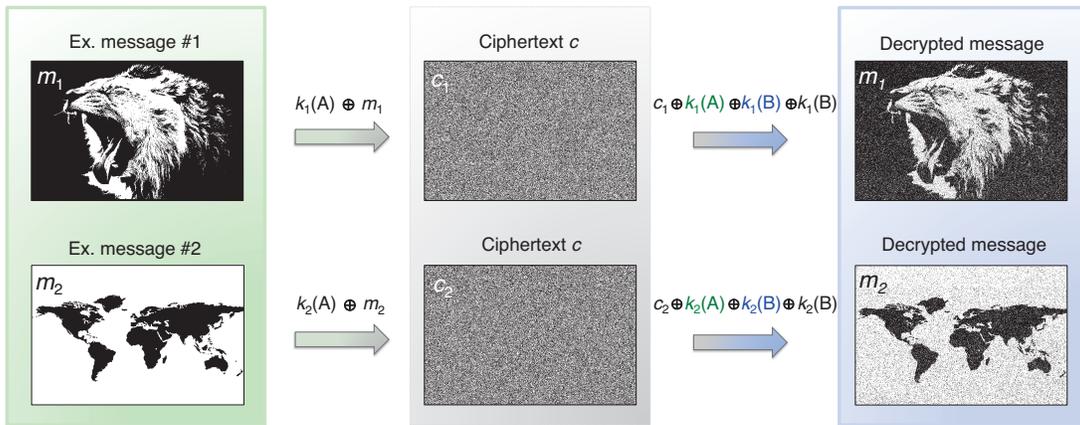}}
\vspace{-.2cm}
\caption{\textbf{Experimental message transmission.} Example messages $m_1$ and $m_2$ are sent between two synchronized CPUF devices 24 hours after dictionary setup. Each message is encrypted to and decrypted from a statistically random ciphertext $c$.}
\label{fig4}
\end{figure*}

The security of the above protocol relies upon the CPUF key sets following what Shannon defines a purely random process~\cite{1}. Possible deviations from pure randomness fall into three categories: correlated bits within the same key, correlations between keys, and the introduction of noise between keys generated at time $t_0$ and at time $t_c$. The sparse projection operator $W$ overcomes such deviations to create keys that asymptotically approach information-theoretic security by sacrificing an increasing number of available encryption bits~\cite{30}. In practice, $W$'s bit reduction factor is selected such that each CPUF key set $k_{1..n}$, viewed as one multi-gigabit random sequence, passes all tests contained within two statistical random number generator test suites commonly accepted as the standards for random sequence certification (the Diehard~\cite{31} and NIST~\cite{32} tests, used often to verify physically generated randomness~\cite{7,8}, are detailed in Supplement F). 

The CPUF devices used in experiment each contain a 2 megapixel transmissive phase SLM imaged onto opal-diffusing glass, serving as our highly random scatterer. A 2 cm light guide (to increase the average detected speckle size) connects the scatterer to a 4.9 megapixel CMOS detector. During public dictionary setup, we display $n=5,000$ different random phase patterns $p$ to generate 174.4 gigabits of raw speckle data from two CPUFs, which is reduced to $N=10$ gigabits of statistically verified~\cite{31,32} randomness via the sparse matrix operator $W$. The approximate theoretical limit of 150 gigabits of randomness per CPUF link (derived in Supplement A) may be achieved using a thicker volumetric scatterer, currently limited to $0.5$ mm for optical stability purposes. Improved-resolution SLM and CMOS arrays may allow random bit densities to eventually reach 1 terabit/mm$^{3}$ (Supplement B).

Experimental communication between two CPUFs following our physically secure OTP protocol is demonstrated $t_{c}=24$ hours after public dictionary setup in Fig.~\ref{fig4}. Due to the slight drift of scatterers, message noise is introduced upon decryption. Error correction helps remove this noise, but reduces the total number of securely transmittable bits by a fixed fraction~\cite{33}. In the included experiment, repetition coding with a code rate of .025 improves the average bit rate error from 0.40 bits to 0.21 bits computed over 100 transmitted 0.4 megabit example messages. 

With a stolen device, an eavesdropper Eve will require approximately 50 hours to characterize the random structure of an ideally operating CPUF at the current capture rate of ~1.5 seconds per key. Faster capture is not possible due to induced scatterer heating (see Supplement G). Coupled with an appropriate device monitoring process, this delay should be sufficient to identify any attempted theft, but currently sets the upper bound on our device's functional lifetime. Additional layers of security, such as encoding shared SLM patterns or enforcing a two-cycle communication requirement, help prevent Eve from obtaining prior communicated keys or utilizing a permanently stolen CPUF. Details of such protocols, as well as a thorough list of possible alternative ``side-channel" attacks, is presented in Supplement G. Our physically secure protocol also makes it impossible for Eve to recover any message after destroying each CPUF, simply achieved by slightly moving or heating the scattering material, which also resets the devices for a new communication round. While information-theoretic security is a good starting point for any new security mechanism, the proposed protocol still currently requires mixing of a message-length key over a secure connection during setup. A public-key protocol adopted to physical CPUF keys can remove any secure connection requirement and significantly reduce required key length, but at the expense of sacrificing perfect OTP security (Supplement H).

In conclusion, the demonstrated CPUF system applies optical scattering to access billions of bits of randomness stored within an unclonable volumetric structure. Information-theoretically secure communication is achieved using a modified OTP protocol. Compared with a large, electronically saved one-time pad, the CPUF's key is extremely challenging to copy or model and can easily scale to provide terabits of repeatable randomness within a small volume. Embedding the device's digital electronics within its volumetric scattering material will further impede any attempted copy or probe attack. With additional study, we hope the convenient properties of optical scattering can solve enough of the OTP's practical shortcomings to rejuvenate interest in its unbreakable security, even in the presence of infinite computing resources.

\section*{Methods}

\subsection*{Methods summary}

One CPUF device uses a spatially filtered and collimated solid-state 532 nm CW laser (Spectra-Physics, Excelsior Scientific 200 mW) to illuminate a transmissive SLM (1920x1080 pixel, 1x1.6 cm Epson HDTV LCD, BBS Bildsysteme). The SLM is operated as a phase modulator (without a second polarizer). The scatterer-detector CPUF segment is composed of four main components that are fastened together using an epoxy to minimize any movement with respect to one another. First, the base of the CPUF is a 2.2 $\mu$m, 2592x1944 CMOS pixel array (The Imaging Source, Micron CMOS MT9P031) with USB readout to a desktop computer. Second, a glass light guide (1.24 cm Quartz disk, Mcmaster-Carr 1357T62) is fixed to the surface of the CMOS protective glass. Third, a custom-printed amplitude-modulating mask (Kodak LVT-exposed on film at 2032dpi, Bowhaus Printing, 5 mm x 5 mm) is attached to the glass light guide to serve as the speckle-shaping aperture. The aperture size is designed to be approximately 1 mm across, ensuring the average speckle size extends across 5 sensor pixels, enhancing speckle image stability.  The apodizing mask follows a 2D-separable Cauchy distribution to ensure the speckle exhibits the required properties of a Markov random process (see Supplement C). 99\% of the mask's transmitted light is contained within its central 1.2 mm$^{2}$ area. Fourth, the volumetric scattering material Ð opal diffusing glass with 0.5 mm scattering volume thickness (Edmund Optics NT46-645) Ð is fixed above the aperture mask. 

\subsection*{Key acquisition and processing}

A low laser power (~0.2 $\mu$W) was used to illuminate each CPUF, preventing speckle pattern decorrelation from material heating but requiring a ~1.5 second image exposure time. Image capture and SLM screen control were driven through a Matlab interface. After capture, the speckle was transformed into a 1D vector and whitened into a key via multiplication with a matrix $W$ (details in Supplement D), with one large sparse $W$ stored locally on a desktop computer for use by two CPUF devices, along with the public dictionary containing the shared set of $n$ binary random SLM patterns and the XOR key-mixtures from connection setup. $W$ need not be unique for each CPUF device, nor kept secret. Each binary SLM pattern was selected uniformly at random from [0, $\pi$] to minimize the probability of key collision. Communication was experimentally achieved between two similarly constructed CPUF devices by populating a public dictionary, waiting 24 hours, and then using the same optical setup to execute the protocol outlined in Fig. 3. Specific encrypt-decrypt parameters for the transmitted messages in Fig. 3(d) are in Supplementary Table 1, where each message contains 0.4 megabits after error correction, as detailed in Supplement E.

\section*{Acknowledgements}

The authors thank Ying Min Wang for constructive discussions, as well as Mark Harfouche and Richard Chen for helpful manuscript feedback. R.H. acknowledges support in part by the National Defense Science and Engineering Graduate (NDSEG) Fellowship Program. B.J. is recipient of the Sir Henry Wellcome Fellowship by the Wellcome Trust.

\section*{Author Contributions}

R.H., B.J. and I.V. and C.Y. conceived and developed the initial idea together. I.V. conceived the final protocol. R.H. designed the experiment, built the setup, collected the data, and performed data and security analysis. All authors collectively wrote the manuscript.

\end{document}